# VR interaction for efficient virtual manufacturing: mini map for multi-user VR navigation platform

Huizhong Cao[a,1], Henrik Söderlund[a], Mélanie Despeisse[a], Francisco Garcia Rivera[b], and Björn Johansson[a]
[a] *Chalmers University of Technology*
[b] *The University of Skövde*
ORCiD ID: Huizhong Cao https://orcid.org/0009-0005-3522-7971

**Abstract.**

Over the past decade, the value and potential of VR applications in manufacturing have gained significant attention in accordance with the rise of Industry 4.0 and beyond. Its efficacy in layout planning, virtual design reviews, and operator training has been well-established in previous studies. However, many functional requirements and interaction parameters of VR for manufacturing remain ambiguously defined. One area awaiting exploration is spatial recognition and learning, crucial for understanding navigation within the virtual manufacturing system and processing spatial data. This is particularly vital in multi-user VR applications where participants' spatial awareness in the virtual realm significantly influences the efficiency of meetings and design reviews. This paper investigates the interaction parameters of multi-user VR, focusing on interactive positioning maps for virtual factory layout planning and exploring the user interaction design of digital maps as navigation aid. A literature study was conducted in order to establish frequently used technics and interactive maps from the VR gaming industry. Multiple demonstrators of different interactive maps provide a comprehensive A/B test which were implemented into a VR multi-user platform using the Unity game engine. Five different prototypes of interactive maps were tested, evaluated and graded by the 20 participants and 40 validated data streams collected. The most efficient interaction design of interactive maps is thus analyzed and discussed in the study.

---

[1] Corresponding Author: Huizhong Cao, huizhong@chalmers.se.

# 1. Introduction

The emerging field of virtual manufacturing requires understanding of virtual user experience and interaction to overcome challenges preventing its adaptation. One of these challenges lies in effectively navigating and understanding the spatial surroundings in the virtual world, a key aspect to successfully completing tasks in a virtual environment [1][2]. The research is focused on XR interaction within a manufacturing context, and by exploring the educational function of gamification, the study will conduct an A/B test on a mini-interactive map (mini-map) with a criterion of NASA-TLX and verify an effective interaction UI for navigation to achieve the optimized educational value of XR applications in manufacturing.

*1.1. XR interaction for manufacturing*

Extended Reality (XR) has been increasingly acknowledged as an advanced tool in manufacturing processes to enable human-centric digital twins beyond Industry 4.0. Industry 4.0 can be defined as the integration of intelligent digital technologies into manufacturing and industrial processes. It encompasses a set of technologies that include industrial IoT networks, AI, XR, Big Data, robotics, and automation. This includes but is not limited to enhanced different use cases enabled by virtual reality (VR) and augmented reality (AR) for manufacturing. E.g., for layout planning, design reviews, virtual prototyping, data monitoring, human-machine interaction and cognitive support [3]. Navigating in VR, e.g., for factory planning, lacks real-world sensory cues, making spatial perception and recognition challenging [4]. This absence can lead to misjudged distances and misunderstandings, especially during multi-user VR interactions [4]. These challenges are critical as they impact not just user experience but also operational efficiency in virtual settings. Users often encounter difficulties in spatial learning and recognition, hampering the efficiency and collaborative potential of the multi-user VR interaction tool [2]. To augment the multi-user experience, this research explores the use of small interactive maps (hereafter called mini maps) with specific user interface (UI) to facilitate spatial recognition, navigation and cooperation.

*1.2. Gamification experience shift*

A pivotal aspect worth exploring is the educational value of games (video, AR, VR) in fostering a robust understanding of mini map practices and related fields. While the educational dimensions of gaming are well-documented [5][6], the focus on mini map for virtual use cases remains relatively unexplored. Furthermore, the interpretation of spatial phenomena through head-mounted displays (HMD) and mini maps in games presents a rich avenue for investigation [7]. The scope extends to studying mini map pragmatism within games, dissecting the interplay between interface interactions and gameplay, and designing diverse game interfaces in a mini map context (non-diegetic, diegetic, meta, spatial) [8][9].

*1.3. NASA TLX & RTLX*

The NASA Task Load Index (NASA TLX) is a well-established method for evaluating perceived workload in various operational settings [10]. A frequent adjustment to NASA-TLX is either the complete removal of the weighting process or individually analyzing the weighted subscales. This unweighted version is termed Raw TLX (RTLX) and is favored by some due to its simplistic application – the scores are either averaged or weighted to deduce the total workload according to overall research objectives. The results of RTLX show a variety of sensitivity, thus validated as a flexible alternative choice when conducting workload assessment [11].

## 2. Objectives and research questions

The objective of this paper is therefore to assess different mini map user interfaces, commonly found in video game and literature, to optimize the navigation experience and explore potential customization plan for multi-user VR applications within manufacturing. Specifically, this study aims to:

- Investigate the usability of various mini map prototypes through user testing.
- Assess the impact of mini map designs on *cognitive workload* using the RTLX - Task Load Index.
- Analyze the practical implications of integrating mini maps into VR environments for manufacturing, focusing on a use case involving factory layout planning.
- Provide discussion and hypothesis on customization strategies for general user interfaces tailored to different decision making and coordination tasks in multiuser VR environments for industrial use in a manufacturing context.

The dominating aim for the study is to evaluate the best mini-map design to minimize cognitive load in virtual manufacturing navigation.

The paper consists of four main sections: the introduction, methods, results, and discussion. The methods section covers both the methodology and the experimental process design. The results section presents the outcomes of the proposed A/B test and the five prototypes, along with the final survey results based on the NASA-TLX. In the discussion section, the authors interpret the interaction UI design principles derived from the results and outline the study's limitations and future research directions.

## 3. Methods

The methodology adapted to answer the research question stated above is two-fold. Firstly, a literature search was conducted to define the existing and explored spatial navigational aid using mini maps. Five mini map categories with different UI specifications from literature and the gaming industry were found [7][18][21][22][23]. In the second step, the map categories gathered from the literature study were compared using A/B testing in an experiment set up where quantitative data about user performance was collected, and a qualitative post experimental survey was documented stemming from responses by each participant in the experiment.

### 3.1. Literature study and gamification exploration

In the initial phase, a literature study and exploration has been carried out using academic databases *Scopus* and *Google Scholar*. The state of art of 3D game industry was explored as complement categories of mini maps. The aim was to identify key research papers, articles, and existing studies focused on VR gaming, mini maps, and UI designs for VR interaction. By extracting and categorizing information, five categories of mini maps were identified based on their UI specifications [7][18][21][22][23]. This literature study served as the theoretical foundation for the subsequent experiment design.

### 3.2. Concept development in VR multi-user platform

Using the *Unity 3D Game Engine,* (Unity Technologies, San Francisco, USA) [12] and the available *Photon Networking* (Exit Games GmbH, Hamburg, Germany) [13] plug-in for *Unity*, a multi-user virtual environment was created allowing for multiple participants to collaborate in a shared digital environment. Furthermore, the *VR Builder* (MindPort GmbH, Siegen, Germany) [14] plug-in to Unity was utilized enabling VR interactions and VR task sequencing in the shared virtual environment. To facilitate collaboration and interaction between players in the shared virtual world life size virtual avatars was added. A virtual model of a factory was introduced creating immersion of the manufacturing context as well as providing a complex layout for the participants to navigate within. Finally, the concepts of mini maps were developed and added to multi-user VR platform as the subject of navigational aid to assess within the VR environment.

### 3.3. User experience investigation through A/B test

A/B testing is a user experience research method that involves a randomized experiment with two variations, A and B. The approach can also extend to multiple variations of a single variable [16]. In the context of this study, A/B testing was used to compare the five different UI specifications of mini maps. Participants were exposed to three types randomly selected mini maps and their performance and user experience were investigated using the survey based on the RTLX index.

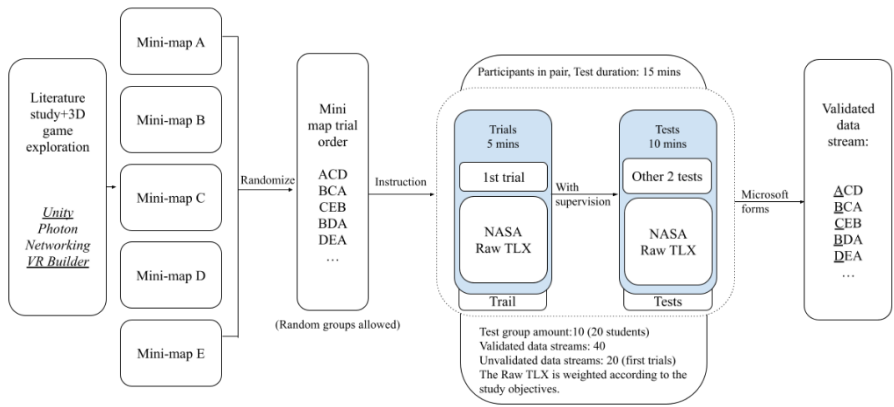

**Figure 1.** The framework of the mini map A/B test is based on simplified RTLX.

For the purpose of this research, a controlled experiment was conducted involving 20 participants, all of them master students from Production Engineering majoring at Chalmers University of Technology. Specific test procedures and game mechanisms are detailed in Figure 1. To ensure reliability and validity, the RTLX index was used to quantitatively evaluate the user experience for each mini map prototype. This helps to correlate the utility of each mini map design with cognitive workload, thereby providing actionable data for further study.

*3.4. Investigation survey*

The survey focused on user experience and perceptions related to the use of the designed VR mini maps in a specific setting of an industrial factory environment.

The survey structure combined both quantitative (ratings) and qualitative (open-ended questions) data. The combination of qualitative and quantitative data provided a broader view of the user experience. The survey was focused on both cognitive load index and specific utilization purposes.

The use of the *RTLX* index [11] provided a standardized method to evaluate the workload experienced by users in terms of:
- Mental demand
  - How mentally demanding was the mini map task?
- Physical demand
  - How physically demanding was the mini map task?
- Temporal demand
  - How hurried or rushed was the pace of the mini map task?
- Effort
  - How hard did you have to work to accomplish your level of performance?
- Performance
  - How satisfied and happy were you with your performance?
- Frustration levels
  - How hard did you have to work to accomplish your level of performance?

In the RTLX assessment, indexes were weighted based on the study objectives. Mental demand and frustration level were primary factors, each weighing 0.3. Secondary importance was attributed to performance and participant satisfaction, each given a weight of 0.2. Temporal demand and effort were also weighted at 0.2. Physical demand, in the context of the gamified session, was deemed irrelevant and assigned a weight of 0. The total weight for all indexes was 1.

## 4. Results

The resulting development and design of the experiment in the VR multi-user platform and the conducted A/B test of the five mini map concepts is explained in further detail in the following sub-sections.

## 4.1. Mini map development

The exploration study of potential mini map concepts resulted in the categorization of five different UI concepts. All the concepts are unique in its design and provides various levels of interactions and capabilities. Making their ability to affect spatial learning and ability to navigate in a virtual space and subject for this study. Using the Unity Game Engine and the multi-user VR platform, as explained in chapter 3, the defined mini map concepts were implemented in a software used to run the experiment. Said experimental software allowed users to experience the different mini map concepts, applied on the same virtual factory, as part of the multi-user VR platform. Table 1 outlines the different mini map categories that were implemented in the software. The last row of the table shows how the different mini maps were implemented in the experimental software.

**Table 1.** Table of the different mini map concepts and their unique descriptions as per the result of authors categorized during the mini map concept exploration.

| Mini Map | A | B | C | D | E |
|---|---|---|---|---|---|
| Description | Hand-held map attached to the VR controller | Hand-held "dollhouse" attached to the VR controller | Static 2D GUI map in the corner of the FOV | In world navigational objects | Toggleable UI window map |
| Detailed description | 2D top view map displayed on an 3D object in the virtual world attached to the VR controller (e.g., mobile device, paper map). | A 3D miniature version of the facility or environment (i.e., dollhouse) attached to the VR controller. | Static 2D top view map attached to the peripheral or corner of the display. Frequently used in desktop gaming applications. | Navigational objects and maps added as static objects in the VR environment (e.g., 3D signs or site maps). | Toggleable UI window in front of the user containing a 2D top view map. Commonly used menu interface in various VR applications. |
| Reference | [7][4][19][21] | [7][4][20] | [17][18][19][11] | [22] | [17][23] |
| Example | 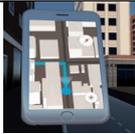 [21] | 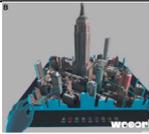 [7] | 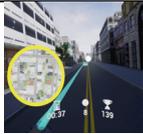 [18] | 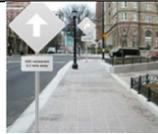 [22] | 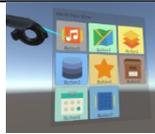 [23] |
| Developed demonstrator with UI specification | 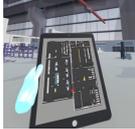 Untogglable 3D + 2D Flexible Small focused zone | 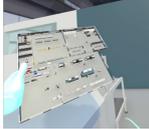 Untogglable 3D Flexible Small focused zone | 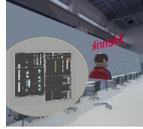 Untogglable 2D Static Medium focused zone | 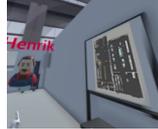 Untogglable 2D Static Medium focused zone | 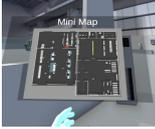 Togglable 2D Flexible Large focused zone |

## 4.2. Gamification setting

In order to be able to evaluate the different mini map categories a small, gamified experience was created within the multi-user VR platform. Using the different versions of the mini maps, the participants were divided into pairs, with each pair engaging in a

VR session featuring the *Over-Maintenance* puzzle game inspired by *Overcooked* (Ghost Town Games, UK) [15], a cooperative game where players collaboratively tackle tasks under time constraints. They were asked to cooperate and carry out a series of tasks. The paired participants were assigned one role each, *warehouse worker* or *machine operator*.

The *warehouse worker* was asked to locate and navigate to a specific shelf in the warehouse of the virtual factory, highlighted on the mini map. Once at the correct shelf, a spare part of a machine should be picked up and taken to the *machine operator,* at the other end of the factory, to be installed on the right machine. In order for the two to be able to arrange for a hand-over and meeting in the VR environment, the two had to rely on the mini map to find each other in the virtual space. The *machine operator* then once more had to rely on mini maps to correctly locate the machine to repair and carry out the task. Once completing the task, using one of the mini map concepts, the experiment was repeated with another mini map and a new layout of the tasks to carry out, after a short pause and survey, making the different mini map concepts subjected to assessment and benchmark. Figure 2 shows an illustration of one of the layouts of the tasks and highlighted objectives, on one of the mini maps, used in the experiment.

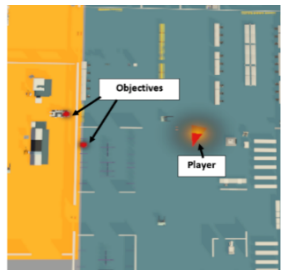

**Figure2.** Snippet of one of the mini maps indicating the players position and the location of the spare part and machine indicated.

### 4.3. A/B Test Results

From the result, the lower the value of the mean RTLX score assessed, the lower workload will be demanded when the specific type of mini map is utilized. Thus, from figure 3, we could see *mini map A – Handheld map* shows the lowest workload demand among five key indexes of RTLX assessment.

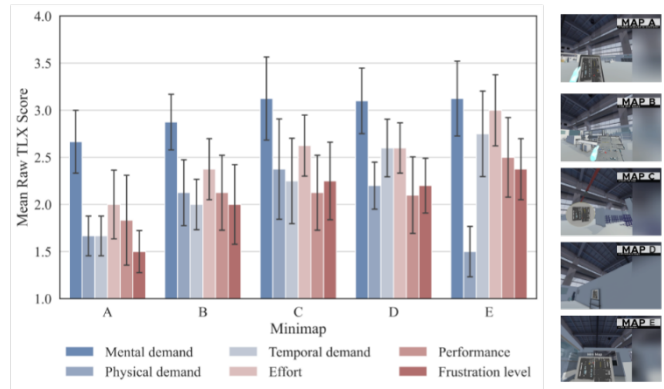

**Figure 3.** Mean RTLX score from 2 tests data and the calculated standard error of measurement (SEM) for each mini map.

The following figure 4 shows the most important indicators according to the research objective – cognitive load assessment, which are mental demand and frustration level.

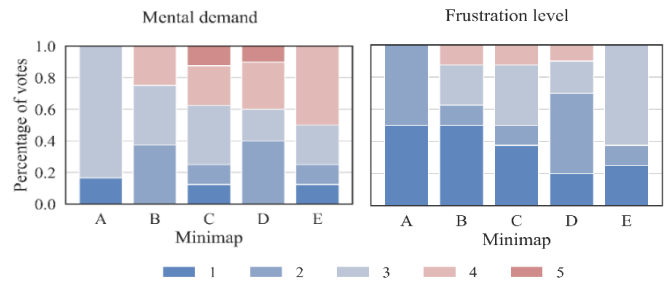

**Figure 4.** User vote distribution of two key indexes: mental demand and frustration level from 2 tests data.

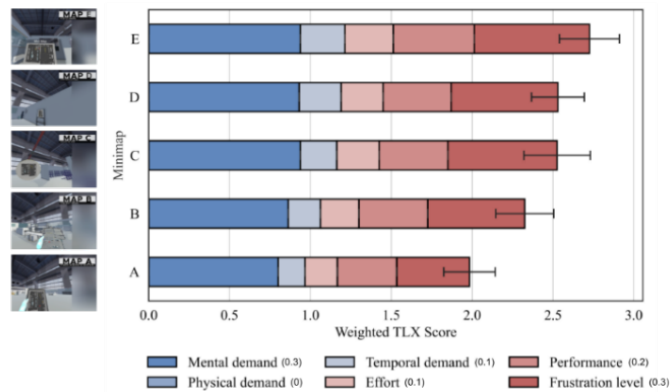

**Figure 5.** Weighted RTLX score of each mini map. The contribution from each category and the calculated total standard error of measurement (SEM) is illustrated from 2 tests data.

Figure 5 shows the RTLX indicators weighted per study objectives. Each mini map illustrates the weighted average score of these indicators.

# 5. Discussion

## 5.1. Reflection on the result

*What type of mini map is the best design for navigation to lowest cognitive load in virtual manufacturing?*

The data from the provided figures (3,4,5) shows that mini map A excels in minimizing workload, as gauged by the RTLX index. Map A had a UI that resembles reality a hand-held tablet which can cause it to be more intuitive. This suggests that real-world mental models might influence the perception of intuitiveness of the mini map. While the study primarily involved VR novices, implementing real-world interfaces can enhance the experience for experts as well. We have concluded two observations:

- *Centralized and focused information dashboard make it easier for core object recognition.*

The first observation posits that the handheld characteristics of mini maps A and B contribute to their reduced cognitive demands. This is potentially due to the malleable positioning of these maps, delivering centralized and focused information. This configuration may enhance the human visual system's efficiency, especially as research indicates superior *core object recognition* [24] within our central visual field, vital for swift object identification during dynamic activities and instant decision-making.

- *The combination of different mini maps UI make it more multi-functional and adaptable.*

The second observation, derived from the findings and interviews, suggests that integrating features from other maps into mini map A might not necessarily reduce the overall workload. Still, it could optimize specific tasks, such as layout perception. Interview feedback indicated that participants, by recalling details from the whiteboard, achieved a heightened understanding of the virtual environment. Potential integrations include:
  - Mini map A+B: Merging A and B could emphasize key objects, like a 3D shelf, on a primarily 2D tablet surface. Retaining lesser important elements in 2D, as in mini map A, might diminish the cognitive effort required for real-time data processing.
  - Mini map A+D: Blending 3D and 2D data could clarify information representation. If the tablet, akin to mini map D's whiteboard, were positioned across various factory spots, it might enhance layout perception and memory cues.
  - Mini map A+E: By amalgamating A with E, a tagging feature can be introduced, enabling users to toggle the mini map's visibility.

## 5.2. Limitation discussion

These hypotheses stem from an initial assessment of mini map prototype designs. Future studies, incorporating a broader participant pool and a comprehensive NASA TLX with Tally [10], are potential for improvement of the data collection.

The experimental framework depicted in Figure 7 utilizes a modified NASA TLX approach for data optimization. Rather than the RTLX used for this study, a more comprehensive and detailed plan could be conducted based on NASA TLX with the weight process by every participant to assess a more individual workload. The modified design also suggests additional validation methods such as error detection for navigation and memorization tests for layout perception. The entire process emphasizes collaboration performance without supervision, evaluated based on task duration.

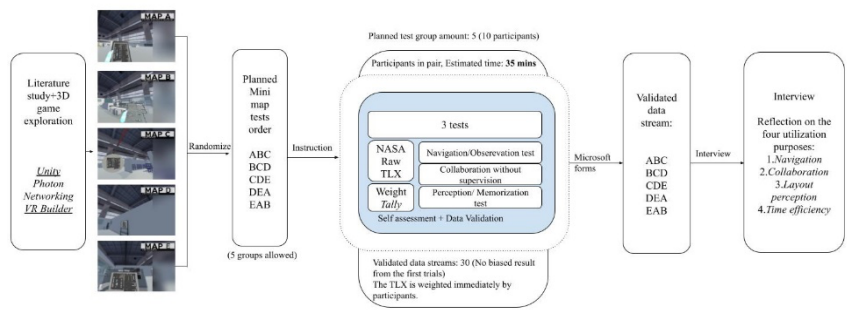

**Figure 7.** Framework of the modified mini map experiment design based on complex NASA TLX for data quality optimization.

*5.3. Future direction*

The future research is based on the resolving the limitation of the experiment design, and also provide new opportunities for mini map development which are:
- Interviews and experiments with industry professionals to gain insights into mini map development.
- Integration of different mini maps to have multifunctional designs which adapt to the four key utilization purposes: layout perception, navigation, collaboration, and time efficiency.
- How to transfer mini maps as things of information and provide more cognitive assistance functions like instruction dashboards integrated into the mini map UI.

**6. Conclusion**

The study shows that the design of the UI in mini maps can affect cognitive workloads for users in multi-user VR environments. This conclusion is based on qualitative and quantitative data gathered from 20 participants, involving 40 validated data streams and five mini-map prototypes. Specifically, Mini-map A, which uses a centralized information board and aligns with users' real-world mental models, demonstrated the lowest cognitive workload. This was measured using key indicators from the RTLX scale and showed a high level of statistical confidence. With the two observations discussed, a further study which will combines features of mini maps for iteration design, will be conducted for larger group of industrial professional participants, and interviews will be conducted as part of user investigation to gain more qualitative insights into the mini map study to reveal the information design UI guidelines for cognitive assistance in virtual environment.

## 7. Acknowledgement

The authors would like to thank the Swedish innovation agency Vinnova for their funding of the PLENUM project, grant number: 2022-01704. The work was carried out within Chalmers' Area of Advance Production. The support is gratefully acknowledged.